\documentclass[epj]{svjour}
\usepackage{graphics}
\makeatletter
\AtBeginDocument{\@ifpackageloaded{natbib}{\ifNAT@numbers\if@filesw\immediate\write\@auxout{\string\global\string
\NAT@numberstrue}\fi\fi}{}}
\makeatother
\usepackage{amstext}
\usepackage{amsmath}

\begin{document}

\title{Comparative study of non-invasive force and stress inference methods in tissue}
\author{
  S. Ishihara,\inst{1,}\thanks{e-mail: shuji@complex.c.u-tokyo.jp}
  K. Sugimura,\inst{2,3}
  S.J. Cox,\inst{4}
  I. Bonnet,\inst{5,6}
  Y. Bella\"{i}che,\inst{5}
  \and
  F. Graner\inst{5,7}
} 
%
%
\institute{
  Graduate School of Arts and Sciences, The University of Tokyo, Komaba 3-8-1, Meguro-ku, Tokyo, Japan
  \and
  Institute for Integrated Cell-Material Sciences (WPI-iCeMS), Kyoto University, Yoshida Honmachi, Sakyo-ku,
Kyoto, Japan
  \and
  RIKEN Brain Science Institute, Wako, Saitama 351-0198, Japan
  \and
  Institute of Mathematics and Physics, Aberystwyth University, Ceredigion, SY23 3BZ, UK
  \and
  Genetics and Developmental Biology, Team ``Polarity, division and morphogenesis,'' Institut Curie, UMR3215
CNRS, U934 Inserm, France
  \and
  Physico-Chimie Curie, Institut Curie, UMR168 CNRS, UPMC, 26 rue d'Ulm, F-75248 Paris Cedex 05, France
  \and
  Laboratoire Mati\`ere et Syst\`emes Complexes, 10 rue Alice Domon et L\'eonie Duquet, F-75205 Paris
Cedex 13, France
}
\date{Received: date / Revised version: date}
%

\abstract{ In the course of animal development, the shape of  tissue emerges in part from mechanical and biochemical interactions between cells. Measuring stress in  tissue is essential for studying morphogenesis and its physical constraints. Experimental measurements of stress reported thus far have been  invasive, indirect, or local. One  theoretical approach is force inference from cell shapes and connectivity, which is non-invasive, can provide a space-time map of stress and relies on prefactors. Here, to validate force-inference methods, we performed a comparative study of them. Three force-inference methods, which differ in their approach of treating indefiniteness in an inverse problem between cell shapes and forces, were tested by using two artificial and two experimental data sets. Our results using different datasets consistently indicate that our Bayesian force inference, by which cell-junction tensions and cell pressures are simultaneously estimated, performs best in terms of accuracy and robustness. Moreover, by measuring the stress anisotropy and relaxation, we cross-validated the force inference and the global annular ablation of tissue, each of which relies on different prefactors. A practical choice of force-inference methods in distinct systems of interest is 
discussed.
} 

\authorrunning{Ishihara \textit{et al}.}
\titlerunning{Force and stress inference in tissue}
\maketitle

%
%

\section{Introduction}\label{sec:intro}
During tissue morphogenesis, cell-level dynamics, \textit{e.g.}, cell morphogenesis, cell rearrangement, cell division, and cell death, are orchestrated in time and space to shape the animal body. As conserved families of signaling pathways in morphogenetic processes have been identified, new challenges arise, such as how mechanical forces that directly act on cells and modify their shapes are integrated with biochemical signaling to regulate the correct tissue patterning  \cite{Lecuit2007,Lecuit2011,Bilder2012,Eaton2012,Kasza2011,Nahmad2011,Wartlick2011}. Indeed, a growing number of studies now address the mechanical basis of morphogenesis. For instance, for epithelial tissue, in which the acto-myosin cytoskeleton is connected to a network of cell-cell junctions (Fig. \ref{fig:Intro}(a)), studies are beginning to clarify how tissue is shaped by forces acting along the plane of the adherens junction, \textit{i.e.}, tension that shortens a cell contact surface and pressure that counteracts the tension to maintain the size of a cell (Fig. \ref{fig:Intro}(b)) \cite{Mao2011,Aigouy2010,Honda2008,Rauzi2008,Farhadifar2007,Graner1993,Hilgenfeldt2008,Kafer2007,Ouchi2003,Staple2010}.

Measuring tissue stress  is therefore useful for deepening our understanding of morphogenesis and its physical constraints. Various \textit{in vivo} mechanical measurement methods have been developed; these include elastography \cite{Ophir1996}, photoelasticity \cite{Nienhaus2009}, magnetic micromanipulation \cite{Desprat2008}, tonometry \cite{Fleury2010}, nanoindentation \cite{Peaucelle2011}, and monolayer stress microscopy (MSM) \cite{Tambe2011}. Among them, laser ablation of individual cell junctions is most frequently used as a tool to evaluate the tension acting on a contact surface of epithelial cells \cite{Rauzi2008,Hutson2003}.

Another approach is based on cell shapes \cite{Stein1982,Brodland2010,Chiou2012,Ishihara2012}. If all cells had the same tensions and pressures, all the angles between cell contact surfaces would be $120^{\circ}$. Conversely, deviations from $120^{\circ}$ would yield information on pressures and tensions. If we manage to have the information well posed, we can infer forces and stresses using only segmented images (that is, images wherein the cell contours and vertices have been recognized). Force inference is non-invasive; hence, spatio-temporal dynamics of forces on more than hundreds of cells can be simultaneously estimated, which represents a distinct advantage over currently available experimental methods. Space-time maps of stress obtained by force inference would unveil the novel physical principles required to regulate morphogenesis.

Given the power of  force-inference methods, their validation in multicellular systems merits thorough and careful analyses. In the present study, we performed a comprehensive test of force inference. Three types of force-inference methods, which differ in their approach to treating indefiniteness in the inverse problem between forces and cell shapes, were employed (Sect. \ref{sec:FIM}). The first method (ST) estimates only tensions, and all the cell pressures are assumed to be the same. The second method (SP) estimates only cell pressures under the assumption of uniform tensions. Under these respective assumptions, the first two methods treat overdetermined problems with respect to unknown variables, that is, cell junction tensions and cell pressures. The third method (STP) treats the ill-conditioned problem and simultaneously estimates both tensions and pressures by employing Bayesian statistics with a prior function representing positive tensions \cite{Ishihara2012}.

Tests of these three force-inference methods were performed for two artificial data, which was generated by  numerical simulations. One advantage of using simulated data is that we can check the accuracy of estimation by directly comparing true and estimated values. One of the artificial data is a simulated foam, which constitutes a well-studied model system for disordered cellular materials \cite{Cantat2010}, and the other is a simulated cell population. In addition, we also used experimental data from epithelial tissues. Patterns of estimated forces were compared among different force-inference methods in \textit{Drosophila} pupal wing and notum (Fig. \ref{fig:Intro}(c)). In notum data, we previously introduced an original type of ablation experiment to measure the mechanical state and material properties of a tissue \cite{Bonnet2012}. The global ablation method and the force-inference methods each depend on different prefactors, and thus the comparison between the two methods provides us with an opportunity to cross-validate them. Based on the results of this study, we will discuss a practical choice of force-inference methods for specific purposes.

%
%

\section{Methods}\label{methods}

\subsection{Systems for tests}
Tests of force inference were performed in two artificial systems and two experimental systems. Artificial data are generated by simulations of foam and two-dimensional tissue. The experimental data of \textit{Drosophila} epithelial tissues (wing and notum) used here were reported in our previous studies \cite{Ishihara2012,Bonnet2012}.

\subsubsection{Numerical simulation of foam}\label{sec:Foam_simulation}
To create the cluster of bubbles shown in Fig. \ref{fig:Foam}, we use the Surface Evolver software \cite{Brakke1992} in a mode that describes each bubble-bubble interface as an arc of a circle with uniform tension. The Evolver minimizes the following energy functional using gradient descent:
\begin{equation}
U_{SE} = \Gamma \sum_{\rm interfaces}  l_{ij} + \sum_{\rm cells} p_i \left(  A_i - A_0 \right)
,\end{equation}
where $\Gamma$ is the line tension (set to one here), $l_{ij}$ is the length of the interface separating bubble $i$ from bubble $j$, and the Lagrange multiplier $p_i$, which ensures that the area $A_i$ of each bubble is constrained to an individual target value $A_0$, is the pressure in bubble $i$. The structure is therefore a precise realization of the ideal two-dimensional soap froth \cite{Cantat2010,Weaire1999}.

We start from a polydisperse foam (\textit{i.e.}, a range of cell target areas) with 2000 bubbles, a total area equal to one, and periodic boundary conditions. This foam is relaxed to an energy minimum. The pressure of each bubble and the position of each vertex in the circular sample of 244 bubbles shown are recorded, and since each arc is defined by the positions of the vertices at its ends and its midpoint, we can calculate the center and radius of curvature for each interface.

In the estimation, bubble-bubble contact surfaces are approximated by straight lines, and the validity of the approximation is discussed later (Sect. \ref{sec:curvature_discussion}).

\subsubsection{Numerical simulation of a cell vertex  model}\label{sec:cellvertexmodel}%
A procedure for generating test data in the cell vertex model \cite{Farhadifar2007,Honda1983} is described in \cite{Ishihara2012}. Briefly, the geometry of cells is approximated by polygonal tiles with straight contact surfaces, each of which is specified by the position of vertex $\vec{r}_i = (x_i, y_i)$ and its connections. The change in  cell geometry is determined by relaxing the following potential function with T1 processes (reconnection of cell contact surfaces) allowed \cite{Honda2008,Rauzi2008,Farhadifar2007,Graner1993,Hilgenfeldt2008,Kafer2007,Ouchi2003}:
\begin{eqnarray}
    \label{eq:VCM}
    U(\vec{r}_{i}) =  \sum_i \frac{K}{2}(A_i \!-\! A_0)^2 +\sum_{[ij]}
    \Gamma_{ij} |\vec{r}_{ij}| + \sum_i \frac{\Lambda}{2} L^2_i.~~~~
\end{eqnarray}Here,
$ \vec{r}_{ij} = \vec{r}_i - \vec{r}_j$ indicates the relative positions of the $i$th and $j$th vertices, and thus $|\vec{r}_{ij}|$ represents the length of the contact surface connecting the $i$th and $j$th vertices.  The first term represents the area elasticity of a cell with stiffness modulus $K=100.0$ and natural area $A_0=1.2$. The second term represents the line tension with constant value $\Gamma_{ij}$, the component of the net tension that is independent of the length of the contact surface. The coefficients of line tension, $\Gamma_{ij}$, are randomly selected from a Gaussian distribution with mean $\langle \Gamma_{ij}\rangle= 0.12 \times KA_0^{3/2}$ and variance $\langle \Delta \Gamma^2_{ij}\rangle^{1/2}   = 0.4 \times \langle \Gamma_{ij}\rangle $. The third term represents cortical elasticity, where $L_i$ is the peripheral length of the $i$th cell and the coefficient is set to $\Lambda = 0.04 \times KA_0$.

\subsubsection{\textit{Drosophila} wing}
The image collection and analysis of \textit{Drosophila} pupal wings (Fig. \ref{fig:Intro}(c)) are described in \cite{Ishihara2012}. Briefly, data collected in pupal wings at 23 hours after puparium formation (h APF) were used for testing the force-inference methods in the present analysis. To highlight the shape of the cell at the level of the adherens junction, D$\alpha$catenin-TagRFP was used as a marker of the adherens junction. We segmented images by using a custom-made macro and plug-ins in ImageJ. We manually corrected the skeletonized pattern when necessary.

\subsubsection{\textit{Drosophila} notum}\label{sec:notum_experiment}
For large scale laser ablation, pupae were mounted as described in \cite{Bonnet2012}. Briefly, we used short laser pulses to sever the adherens junctions in an annular region around an approximately 30-$\mu$m-radius circular tissue domain. The wound margin retracted with a speed reflecting the stress prevailing in the tissue before ablation, and its anisotropy. For each experiment presented in \cite{Bonnet2012}, using in-house automatic software (\textsc{MATLAB} \cite{Bosveld2012}) based on a watershed algorithm followed by manual checking, we segmented the cell contours in the circular patch of tissue before and after its periphery was ablated. To increase the image quality, the initial image of the tissue at rest was obtained as an average of 10 images before ablation. Similarly, the final image was obtained as an average of raw images ($n=3$ or $n=11$ according to the quality of the movie) when the ablated circular tissue domain was again at rest.

\subsection{Force-inference methods}\label{sec:ForceInference}
\begin{figure}
  \centering
  \resizebox{0.48\textwidth}{!}{%
    \includegraphics{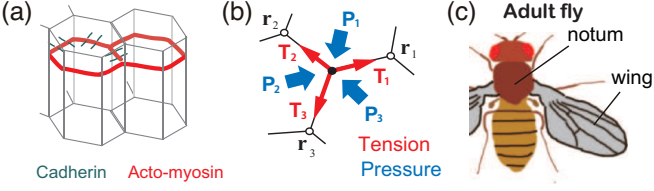}
  }
  \caption{ The structure and force balance of an epithelial tissue. (a) Mechanical interactions among epithelial cells act mostly in the plane of the adherens junction, where cell adhesion molecules, cadherin, held cells together. Inside the cell, an acto-myosin cable runs along the cell cortex in the plane of the adherens junction. (b) Forces acting on a vertex located at position $\vec{r}_0$ indicated by a black dot.  Tensions along the cell contact surfaces pull the vertex in the respective directions indicated by red arrows, while the cell pressures push the vertex in the directions indicated by blue arrows. (c) Two epithelial tissues: wing and notum in \textit{Drosophila}, shown in the adult fly (experiments performed in pupa).  }
  \label{fig:Intro}       
\end{figure}

Here, we briefly outline how one infers forces and stress from patterns of epithelial cell shapes and their connectivity. Detailed descriptions of force inference can be found in \cite{Chiou2012,Ishihara2012}. As input, we take a segmented image of epithelial cells. Epithelial tissue is approximated as a two-dimensional sheet, and cells are represented by polygonal tiles. Because the curvature of the cell contact surface is small in most epithelia, we approximate a cell contact surface as a straight edge. Here, the unknowns are the tension of each contact surface and the pressure of each cell. If the deformation of cells is sufficiently slow in a dissipative environment, these forces are almost balanced, and the system lies in the vicinity of an equilibrium state (quasistatic assumption). Then, by considering force-balance equations with a given cell geometry, one can infer tensions and pressures.

\subsubsection{Equations}
This section presents the force-balance equations used to deduce forces and stress. Although the derivation of the force-balance equations was described in \cite{Ishihara2012}, we reiterated it to make the present paper self-contained. The geometry of the tissue is specified by the positions of vertices, $\vec{r}_i = (x_i,y_i)$, and their connectivity. We denote the tension of the cell contact surface that connects the $i$th and $j$th vertices as $T_{ij}$ and the pressure of the $i$th cell as $P_i$.  Let us consider the force acting on the $0$th vertex at the origin $\vec{r}_0= (0,0)$ in Fig. \ref{fig:Intro}(b). The forces in the $x$ and $y$ directions are given by
\begin{eqnarray}
    F^x_0 &=& \sum_{i=1}^3\frac{ x_{i}}{|\vec{r}_{i}|}T_{i}-\sum_{i}^3\frac{y_i}{2}\left(P_{i}-P_{i+1}\right),
\label{eq:Force_Vertex_X}\\
    F^y_0 &=& \sum_{i=1}^3\frac{ y_{i}}{|\vec{r}_{i}|}T_{i}+\sum_{i}^3\frac{x_i}{2}\left(P_{i}-P_{i+1}\right).
\label{eq:Force_Vertex_Y}
\end{eqnarray}
Here, $T_{i} = T_{i0}$ and $P_4 = P_1$. By using the orientation of the edge, $\theta_{ij} = \tan^{-1} \left(y_{ij}/x_{ij}\right)$, the coefficients of $T_{ij}$ are simply expressed as $x_{ij}/|\vec{r}_{ij}| = \cos \theta_{ij}$ and $y_{ij}/|\vec{r}_{ij}| = \sin \theta_{ij}$. Pressures act all along the sides of the cells in their normal direction, and projection to the $x$ and  $y$ axes  gives the prefactors $-y_i$ and $x_i$, respectively, of the normal force. Half of the force acts on the end-points of the cell contact surface, represented by the second terms in Eqs. (\ref{eq:Force_Vertex_X}) and (\ref{eq:Force_Vertex_Y}).

More rigorously, Eqs. (\ref{eq:Force_Vertex_X}) and (\ref{eq:Force_Vertex_Y}) can be obtained by differentiating of a potential function. Consider a potential function $U(\{\vec{r}_i\})$ that determines the tissue's mechanical characteristics, as exemplified in Eq. (\ref{eq:VCM}). The derivative of the potential function with respect to $\vec{r}_i$ gives the forces on the $i$th vertex as $\vec{F}_i = \partial U(\{\vec{r}_i\})/\partial \vec{r}_i$.  $\vec{F}_i$ can be written as
\begin{eqnarray}
  \vec{F}_i = \sum_{[jk]}\frac{\partial U}{\partial |r_{jk}|}\frac{\partial |r_{jk}|}{ \partial \vec{r}_i}
+
 \sum_j \frac{\partial U}{\partial A_j}\frac{\partial A_j}{\partial \vec{r}_i},
\end{eqnarray}
where $|r_{jk}|$ and $A_j$ are the length of the cell contact surfaces and the cell area, respectively. With the definition of pressure and tension, $P_j \equiv - \partial U/\partial A_j$ and $ T_{jk} \equiv\partial U/\partial |r_{jk}|$, the above equation leads to Eqs. (\ref{eq:Force_Vertex_X}) and (\ref{eq:Force_Vertex_Y}), irrespective of the functional form of $U$.

Suppose we have an image in which $N$ cells are surrounded by $R$ cells. The numbers of cell contact surfaces and vertices in the image are denoted as $E$ and $V$, respectively.  Repeating the same derivation of force-balance equations for every vertex, we obtain a vector $\vec{F} = (\vec{F}^x, \vec{F}^y)$ that represents the forces acting on vertices in the $x$ and $y$ directions as
\begin{eqnarray}
  \label{eq:Force_MatrixForm}
  \vec{F} = A_T \vec{T} + A_P \vec{P} = A \vec{X} .   
\end{eqnarray}
Here, $\vec{T} $ and $\vec{P} $ are vectors composed of $T_{ij}$ and $P_{i}$, respectively.  $\vec{X} = (\vec{T}, \vec{P})$ represents the unknown variables to be inferred. $A_T$ and $A_P$ (and thus $A$) are $2V \times E$ and $ 2V \times (N+R)$ matrices representing the coefficients of $T_{ij}$ and $P_{i}$ in Eqs. (\ref{eq:Force_Vertex_X}) and (\ref{eq:Force_Vertex_Y}), respectively, and they are determined by the positions of the vertices. Under the assumption of quasistatic cell shape changes, the force-balance equation becomes
\begin{eqnarray}
  \label{eq:FB}
  A_T \vec{T} + A_P \vec{P} = 0.      
\end{eqnarray}
Equation (\ref{eq:FB}) gives us a relationship between the observable geometry (angles and lengths; see Eqs. (\ref{eq:Force_Vertex_X}) and (\ref{eq:Force_Vertex_Y})) of cells and the unobservable tensions $\vec{T}$ and pressures $\vec{P}
$.

In Eq. (\ref{eq:FB}), the scale factor of forces is undetermined, because the cell shape does not provide any information about it. The force-inference method therefore estimates relative values of forces in the scale, as described below.  In addition, hydrostatic 
pressure (the baseline value of pressure) cannot be determined, because Eq. (\ref{eq:FB}) is invariant under a uniform increase of pressure (see Eq. (\ref{eq:Force_Vertex_X})). Thus, it is differences in pressures among cells that are inferred. The estimated tensions and pressures are related to true ones as $\vec{T} = c\vec{T}_{true}$ and $\vec{P} = c\vec{P}_{true}+\Delta \vec{p}$. Unless mentioned explicitly, the prefactor $c$ is selected to satisfy the requirement that the average of the tensions should be unity:
\begin{eqnarray}
  \scriptsize{\sum\nolimits}_{[ij]}T_{ij}/E = 1 \label{eq:Tuni}
,\end{eqnarray}
and the hydrostatic pressures are selected such that the average of cell pressures is zero:
\begin{eqnarray}
  \scriptsize{\sum\nolimits}_i P_i = 0. \label{eq:Pzero}
\end{eqnarray}

As described in our previous study \cite{Ishihara2012}, the critical difficulty in calculating forces from Eq. (\ref{eq:FB}) originates from an insufficient number of force-balance equations to determine the unique solution of the unknown variables $\vec{X}$. By considering the topology of a network of cell contact surfaces, we can show that the number of unknowns ($N+R+E$) is smaller than the number of conditions $(2V)$ by $R+f+1$, where $R$ and $f$ are the number of surrounding cells and the number of four-way junctions, respectively. One indefiniteness arises from the hydrostatic pressure as described above, and another $R+f$ indefiniteness results from the boundary conditions and four-way junctions. The latter indefiniteness should be handled by a proper assumption on a system to get plausible estimates.

One possible formulation of the inverse problem is to decrease the number of unknown variables by assuming a relationship among variables (variable-reduction approach). For example, Chiou \textit{et al}. \cite{Chiou2012} set  all cells to have the same pressure $P_i = P_0$. Then, with a constraint of finite scale, given by Eq. (\ref{eq:Tuni}), the inverse problem becomes over-determined and has a unique solution. Another formulation is to adopt Bayesian statistics, which  now provides a standard framework for this ill-conditioned problem, where our expectation for the system is incorporated as a prior (Bayesian approach) \cite{Akaike1980,Kaipio2004}. For example, we used a prior function expecting that tension is distributed around a positive value \cite{Ishihara2012}. In a Bayesian framework, forces are inferred by maximum \textit{a posteriori} (MAP) estimation after the marginal likelihood is maximized with respect to a hyperparameter.

With the obtained values of tensions and pressure ($\vec{X} = (\vec{P},\vec{T})$), one can integrate them to deduce the global stress.  The stress tensor is evaluated with the Batchelor stress tensor given by \cite{Ishihara2012,Batchelor1970}
\begin{eqnarray}
    \label{eq:StressTensor}
    \mbox{\boldmath $\sigma$} = \frac{1}{A}\left( -\sum_i P_i A_i {\bf I}+\sum_{[ij]} T_{ij} \frac{\vec{r}_{ij}\otimes
\vec{r}_{ij}}{|\vec{r}_{ij}|}\right)
,\end{eqnarray}
where ${\bf I}$ is the two-dimensional identity matrix and $A \equiv \sum_i A_i$ is the total tissue area. Because the scale and the hydrostatic pressure are undetermined in the force-balance equations, the scale of {\boldmath  $\sigma$} is also undetermined as well as the additional pressure $-\Delta p {\bf I}$. Some quantities derived from the tensor are independent of $\Delta p$; they include the maximum stress direction and the difference of two eigenvalues of {\boldmath $\sigma$}.

\subsubsection{Force inference methods to be tested}\label{sec:FIM}
In this study, we performed a comparative test on three types of force-inference methods, which differ in their approach of treating indefiniteness in the inverse problem between forces and cell shape.  The first two methods reduce the number of unknown variables and treat overdetermined problems with respect to unknown variables $\vec{T}$ or $\vec{P}$ (variable-reduction approach). The third method treats the ill-conditioned problem by employing Bayesian statistics (Bayesian approach) \cite{Ishihara2012}. We have called them ST, SP and STP, where the ``S" stands for  ``straight" edges (curvatures are neglected and cells are treated as polygons); ``T" and ``P" mean that tensions and pressures are unknown, respectively.

\vspace{4mm}
\noindent{\textit{\textsf{Tension inference under an assumption of uniform pressure (ST)}}}\\
The first method ST estimates only tensions. In ST, the difference in pressures among cells is assumed to be small and cells are approximated to have the same pressure $P_i=P_0$.  Under this assumption,  $A_P \vec{P}_0$ vanishes, as is immediately evident from Eq. (\ref{eq:Force_Vertex_X}) and (\ref{eq:Force_Vertex_Y}).  Then, Eq. (\ref{eq:FB}) becomes
\begin{eqnarray}
  \label{eq:ST}
  A_T\vec{T} = 0.
\end{eqnarray}
This equation is overdetermined and its solution is found by minimizing $|A_T\vec{T}|^2$ with the constraint given by Eq. (\ref{eq:Tuni}). That is, estimation of $\vec{T}$ is given by the eigenvector of the smallest eigenvalue of matrix $A_T^tA_T$ with a normalization factor used to satisfy Eq. (\ref{eq:Tuni}).

\vspace{4mm}
\noindent {\textit{\textsf{Pressure inference under an assumption of uniform tension (SP)}}}\\
The second method SP estimates only cell pressures. Under the assumption that all tensions are uniform, \textit{i.e.}, $T_{ij} =1$, which is strictly exact in the case of foam, Eq. (\ref{eq:FB}) becomes
\begin{eqnarray}
  \label{eq:SP_1}
  A_P\vec{P} = - A_T\vec{T}_0,
\end{eqnarray}
where all the components of $\vec{T}_0$ are $1$. This problem is also overconditioned, and estimates of pressures are found by minimizing $|A_P \vec{P} + A_T\vec{T}_0|^2$ with respect to $\vec{P}$.  The solution is given as
\begin{eqnarray}
  \label{eq:SP}
  \vec{P} = - \tilde{A}_P^{-1} A_T\vec{T}_0,
\end{eqnarray}
where $\tilde{A}_P^{-1}$ is the Moore--Penrose pseudo-inverse matrix of $A_P$. Equation (\ref{eq:SP}) can be shown to satisfy Eq. (\ref{eq:Pzero}).

\vspace{4mm}
\noindent {\textit{\textsf{Bayesian inference of tensions and pressures with a prior of positive tension (STP)}}}\\
The third method STP is a Bayesian inference of tensions and pressures developed by two of us \cite{Ishihara2012}. Briefly, force inference is carried out by MAP estimation, \textit{i.e.}, by taking the maximum value of an \textit{a posteriori} distribution given by
\begin{eqnarray}
  \label{eq:Posterior}
  P(\vec{X}) \propto e^{-|A\vec{p}|^2/2\Sigma^2 } \times \pi(\vec{X}),
\end{eqnarray}
where $\pi(\vec{X})$ is a prior function. We adopted the prior that $T_{ij}$ is distributed around a positive  value, because laser severing experiments indicate that tensions along cell contact surfaces are usually constricting in  epithelial tissue \cite{Rauzi2008,Hutson2003}. Hence,
\begin{eqnarray}
    \pi(\vec{X}) = (2\pi\omega^2)^{-E/2} e^{-{\scriptstyle
        \sum}_{[ij]}(T_{ij}-T_0)^2/2\omega^2} \!\times \delta
    \!\left({\scriptstyle \sum}_i P_i\right).~~~
\end{eqnarray}
A Gaussian distribution of $T_{ij}$ around $T_0 > 0$ represents our expectation explained above, and the Dirac $\delta$ function is introduced to satisfy Eq. (\ref{eq:Pzero}). We can select $T_0 =1$ by adjusting the scale factor.  A criterion to determine hyperparameters $\Sigma^2$ and $\omega^2$ is to minimize the Kullback-Leibler distance between the probability function parameterized by $(\Sigma^2, \omega^2)$ (\textit{i.e.}, statistical model) and empirical distribution. This is formulated by maximizing marginal likelihood, or equivalently, by minimizing the Akaike Bayesian information criterion (ABIC) \cite{Akaike1980,Kaipio2004}.
\begin{eqnarray}
L(\Sigma^2, \omega^2) = \int P(\vec{X}|\Sigma^2,\omega^2)d\vec{X}  . \label{eq:ML}  
\end{eqnarray}
We compared estimation using different priors in the artificial data generated by a cell vertex model and confirmed that estimation using the prior expecting positive tensions gave the best fit with true values of forces. Details are described in \cite{Ishihara2012}.

\subsubsection{Variable-reduction and Bayesian approaches}\label{sec:Variable-reductionandBayes}
It is instructive to mention how variable-reduction and Bayesian approaches are related.  Both  approaches can be formulated as the minimization of the function
\begin{eqnarray}
    S(\vec{X}) = \left|A\vec{X}\right|^2  + \mu  H(\vec{X})
\end{eqnarray}
with respect to $\vec{X} = (\vec{T}, \vec{P})$ with constraints given by Eqs. (\ref{eq:Tuni}) and (\ref{eq:Pzero}).  $H(\vec{X})$ in the second term indicates our expectation of the system, which compensates for indefiniteness in the force-balance equation represented by the first term. Both SP (variable reduction) and STP (Bayesian) are formulated by using $H(\vec{X}) = \sum_{[ij]} (T_{ij}-T_0)^2$, and their difference is the weight $\mu$.

In the variable-reduction approach, one assumes that the expectation given by $H(\vec{X})$ is a strict constraint to be satisfied, and thus the coefficient $\mu$ is given as the Lagrange multiplier. In contrast, in the Bayesian approach, $S(\vec{X})$ is related to the 
posterior probability equation (Eq. (\ref{eq:Posterior})) as $P(\vec{X};\mu,\Sigma^2)$ $\propto \exp \left(-S(\vec{X};\mu)/ 2\Sigma^2\right)$ with $ \mu = \Sigma^2/\omega^2$.  The minimization of $S(\vec{X})$ is equivalent to MAP estimation.  The expectation $H(\vec{X})$ is incorporated into the prior function, and it is not required to be strictly satisfied. The degree of deviation is controlled by $\mu$, the weight of the second term, which is objectively determined by maximizing the marginal likelihood $L(\Sigma^2,\omega^2)$ (Eq. (\ref{eq:ML})). By using this procedure, the first term (fit of data) and the second term (expectation) are balanced by taking into account the data quality \cite{Ishihara2012}.

%
%

\section{Results} \label{results} %

\subsection{Numerical data for foam}

\begin{figure*}
  \centering
  \resizebox{.8\textwidth}{!}{%
    \includegraphics{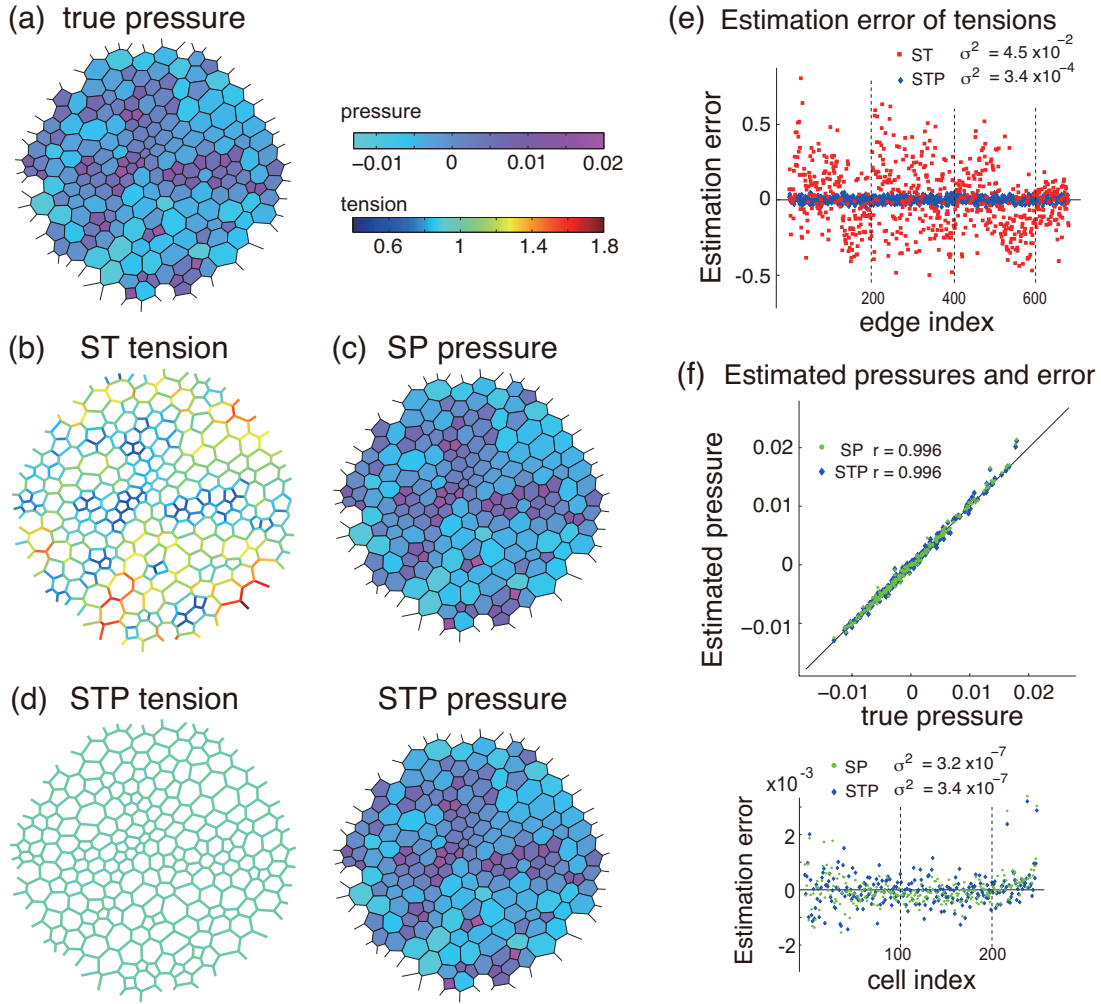}
  }
  \caption{ True and estimated forces for artificial foam data.  (a) A color map of true pressures of the artificial data.  True tensions 
are set to be $T_{ij} = 1$ for all contact surfaces in the foam (not shown). (b) Tensions estimated using ST.  (c) Pressures estimated 
using SP.  (d) Tensions (left) and pressures (right) estimated using STP. (e) Estimation errors of tensions for ST (red) and STP 
(blue).  (f) Estimated pressures plotted against true ones (top) and their errors (bottom).  Pressures estimated using SP and STP 
are indicated with green and blue points, respectively.  }
  \label{fig:Foam}
\end{figure*}

Foam and an epithelial tissue are disordered cellular materials \cite{Kafer2007,Cantat2010,Thomson1917} and they have distinct mechanical natures. Most significantly, the tension of each contact surface is always kept uniform in foam \cite{Cantat2010}.  Here, the geometry of a foam (positions of vertices and their connectivity) is obtained by numerical simulation (see Sect. \ref{sec:Foam_simulation}), and it is provided as an input for force inference. Although numerical simulation is carried out by considering the curvature of the membrane, we approximate it as a straight line when estimating forces and stress. The curvature of the artificial data for foam used here is not significant: the angular error caused by neglecting the curvature is $5.0^{\circ} (\pm 4.9^{\circ})$.

Figure \ref{fig:Foam}(a) shows true bubble pressures  indicated by a color scale. The true tensions of all contact surfaces are $T_{ij} = 1$. We conducted the force-inference by ST, SP, and STP, where the ``S" stands for  ``straight" edges (curvatures are neglected and cells are treated as polygons); ``T" and ``P" mean that tensions and pressures are unknown, respectively (see Sect. \ref{sec:FIM}). The results of force inference by ST, SP, and STP are shown in Figs. \ref{fig:Foam}(b)-(f) and are summarized in Table \ref{tab:ResultSum}. Since the pressures of all bubbles are set to be uniform in ST, only estimated tensions are shown in Fig. \ref{fig:Foam}(b). We notice that contact surfaces belonging to small bubbles were estimated to have smaller tension. The estimation errors of tensions for individual contact surfaces are shown in Fig. \ref{fig:Foam}(e) (red points).  Their deviation from the true values (the mean residue of error) was $\sigma^2 = 4.5 \times 10^{-2}$.

The pressures estimated using SP are indicated with a color map in Fig. \ref{fig:Foam}(c), and they are plotted against the true ones in Fig. \ref{fig:Foam}(f) (green points). The results indicate that the accuracy of force inference was higher in SP than in ST (with the correlation between true and estimated pressures being $r=0.996$ in Fig. \ref{fig:Foam}(f), upper panel, and $\sigma^2 =3.2 \times 10^{-7}$ in Fig. \ref{fig:Foam}(f), bottom panel).

The estimated tensions and pressure obtained using STP are shown in Fig. \ref{fig:Foam}(d), and these estimated values are compared with the true values in Figs. \ref{fig:Foam}(e) and (f) (blue points). The estimation errors of tensions in STP are considerably smaller than those in ST ($\sigma^2 = 3.4 \times 10^{-4}$). In addition, estimated pressures are well correlated with the true ones as in SP ($r=0.996$; Fig. \ref{fig:Foam}(f)). The error in the estimated pressures is very small ($\sigma^2 = 3.4 \times 10^{-7}$; Fig. \ref{fig:Foam}(f), bottom panel) but slightly larger than for SP, where true tensions are set.

\subsection{Numerical data for the cell vertex model}

\begin{figure*}
  \resizebox{.99\textwidth}{!}{%
    \includegraphics{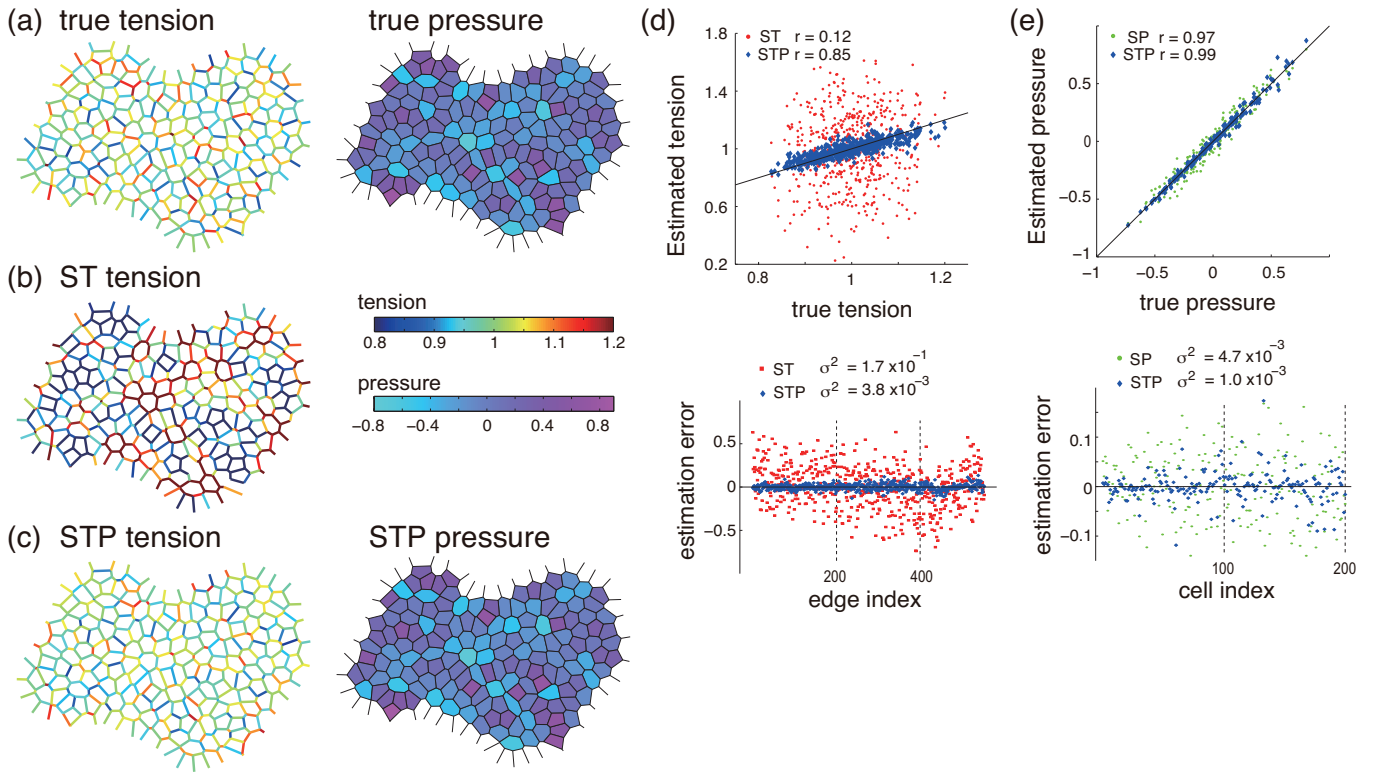}
  }
  \caption{ True and estimated forces in artificial data obtained by cell vertex model. (a) The true tensions (left) and pressures (right) in the artificial data. (b) Tensions estimated using ST.  (c)  Tensions (left) and pressures (right) estimated using STP.  (d) Top: Estimated tensions plotted against true ones for ST (red) and STP (blue). Bottom: Estimation errors of tensions.  (e) Estimated pressures plotted against true ones (top) and their errors (bottom). Pressures estimated using SP and STP are indicated with green and blue points, respectively.  }
  \label{fig:CVM}       
\end{figure*}

To evaluate the force-inference methods, we generate artificial data by simulating a cell vertex model with random parameters (see Sect. \ref{sec:cellvertexmodel}). The results of this test for STP were reported previously \cite{Ishihara2012}.

Figure \ref{fig:CVM}(a) shows maps of true forces, and Figs. \ref{fig:CVM}(b) and (c) show maps of forces inferred using ST and STP, respectively. The pressure map obtained using SP is very similar to that obtained using STP (not shown). In ST, the estimated and true tensions do not correlate well with each other ($r=0.12$), and the deviation of the estimated tensions from the true ones is $\sigma^2 = 0.17$ (red points in Fig. \ref{fig:CVM}(d)). SP provides good estimates of pressures with high correlation $r=0.97$ and small mean residue of error $\sigma^2 = 4.7 \times 10^{-3}$ (green points in Fig. \ref{fig:CVM}(e)). For STP, the correlations with true values of tensions and pressures are $0.85$ and $0.99$, respectively (Fig. \ref{fig:CVM}(d) and (e), blue points).

\subsection{\textit{Drosophila} wing}
\begin{figure*}
  \resizebox{.99\textwidth}{!}{%
    \includegraphics{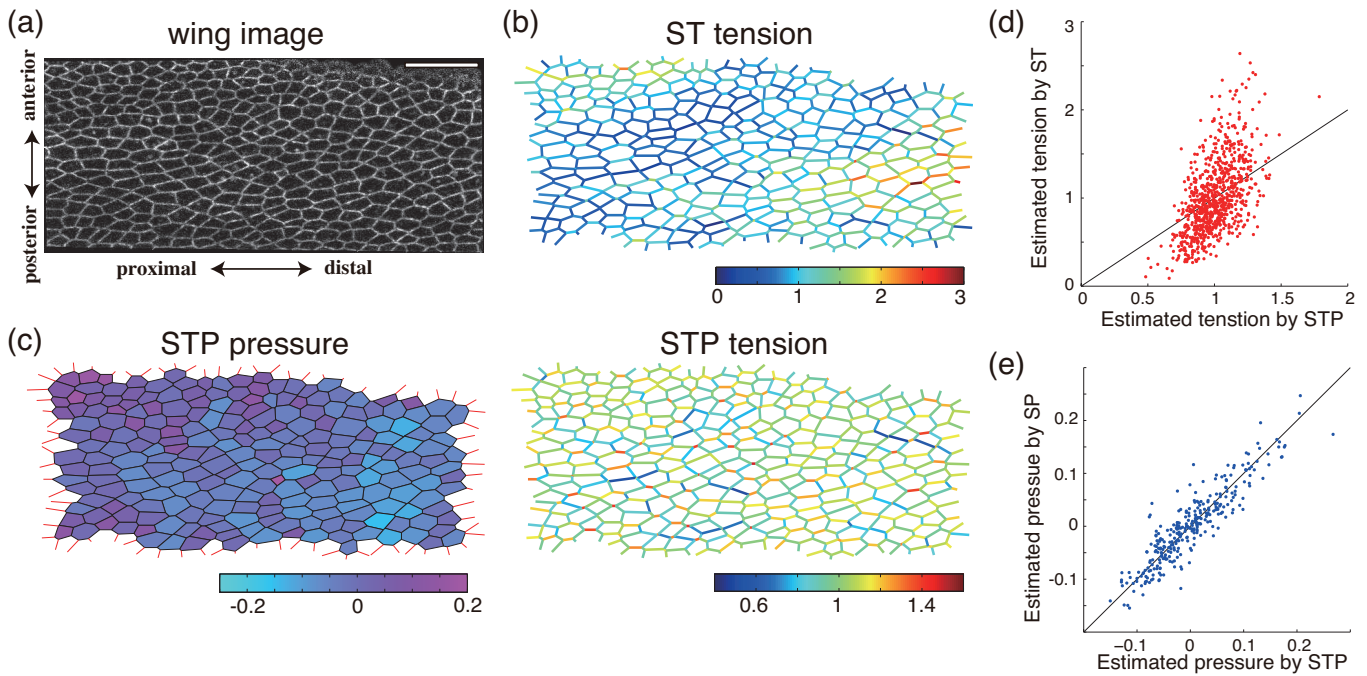}
  }
  \caption{ Estimated tensions and pressures for a \textit{Drosophila} pupal wing.  (a) An image of a \textit{Drosophila} wing at 23 h APF. D$\alpha$catenin-TagRFP is used to highlight cell shape. Scale bar: $20\;\mu \text{m}$. (b) and (c) Color maps of estimated tensions and pressures, respectively. A color scale is shown for each image. (b) Cell-junction tensions estimated using ST. (c) Tensions (top) and pressures (bottom) estimated using STP. (d) Comparison of estimated tensions obtained using STP and ST. (e) Comparison of estimated pressures obtained using STP and SP.}
  \label{fig:wing}       
\end{figure*}

We apply the three force-inference methods to an image (Fig. \ref{fig:wing}(a)) of a \textit{Drosophila} pupal wing (Fig. \ref{fig:Intro}(c)).  The results of estimations obtained using STP, which agree with laser ablation of a contact surface and the myosin distribution, were reported in \cite{Ishihara2012}.

Figure \ref{fig:wing} shows the results of force estimations obtained using ST (b) and STP (c). Estimated pressures obtained by SP exhibit maps similar to those obtained by STP (not shown). Since the true values of tensions or pressures are not known for experimental data, we compared tensions obtained using ST and STP and pressures obtained by SP and STP, respectively.  The pressures obtained using SP and STP show a good correlation as in the artificial data for foam and the cell (Figs. \ref{fig:wing}(c) and (e)); thus pressure maps obtained by the two methods are very similar (the pressure map obtained using SP is not shown).  However, the tensions estimated using ST show a larger deviation than those estimated using STP (Figs. \ref{fig:wing}(b)-(d)); the standard deviation of tensions given by ST is $0.42$, whereas that given by STP is $0.15$ (with similar results being obtained using all samples of the wings examined ($n = 20$)).

\subsection{\textit{Drosophila} notum}

\begin{figure}
  \centering
  \resizebox{.48\textwidth}{!}{%
  \includegraphics{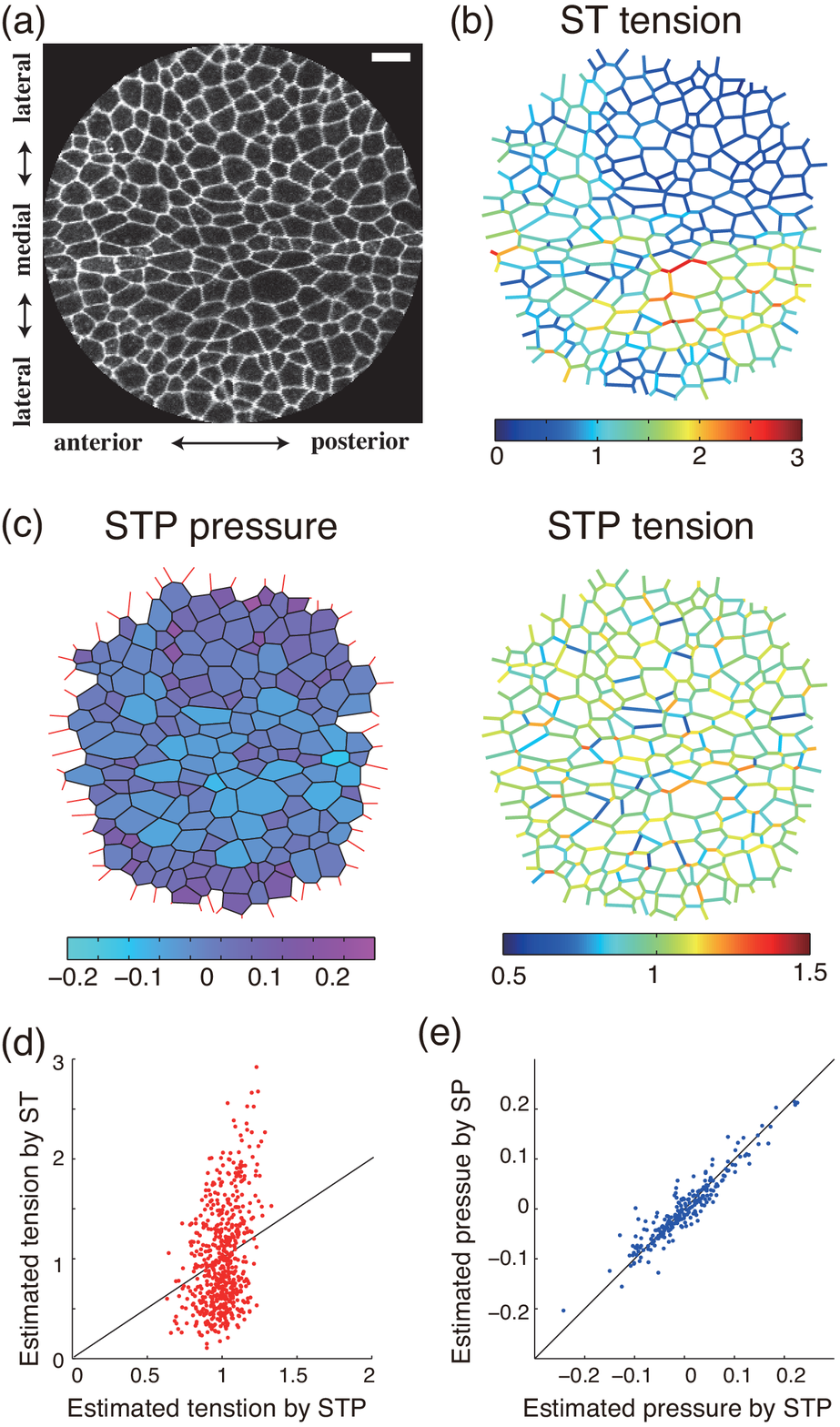}
  }
  \caption{ Estimated tensions and pressures for a \textit{Drosophila} pupal notum. Data from \cite{Bonnet2012} are used. (a) An image of a \textit{Drosophila} notum. Scale bar: $ 10 \;\mu\text{m}$. (b) and (c) Color maps of estimated tensions and pressures. A color scale is shown for each image. (b) Cell-junction tensions estimated using ST. (c) Tensions (top) and pressures (bottom) estimated using STP. (d) Comparison of estimated tensions obtained using STP and ST. (e) Comparison of estimated pressures obtained using STP and SP.}  \label{fig:notum}
\end{figure}

We applied the force-inference methods to the \textit{Drosophila} notum (Fig. \ref{fig:Intro}(c) and Fig. \ref{fig:notum}(a)). The experimental data used were collected at three developmental stages referred to as young (around 18 h APF), middle (around 22 h APF), and old (around 26 hr APF) (see details in \cite{Bonnet2012}).

The results of our analysis on the notum are consistent with those on the other three systems. Pressures obtained using SP and STP show a good correlation (Fig. \ref{fig:notum}(e)), whereas estimated tensions obtained using ST are more disperse than those from STP (Figs. \ref{fig:notum}(b)--(d)); the standard deviations of tension distributions are $0.51$ and $0.11$ for ST and STP, respectively (with similar results being obtained from all samples of the notum examined ($n = 23$)).

\begin{figure}[htbp]
  \centering
  \resizebox{0.48\textwidth}{!}{%
    \includegraphics{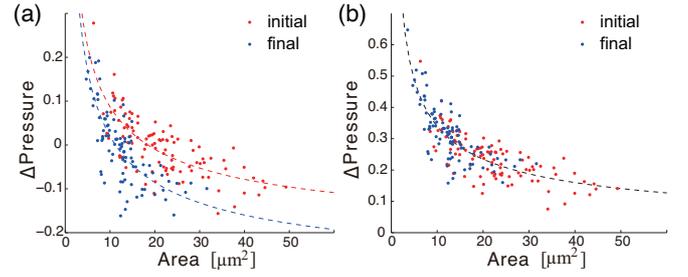}
  }
  \caption{ The calibration of scale factors of the stress tensor.  (a) Estimated values of cell pressures plotted against cell area before cells were ablated (red) and after the  tissue is relaxed (blue). (b) The area--pressure relation is assumed to be maintained between the two time points. By using a fitting function $P(A) = a/\sqrt{A}+b$, the pressures at the initial stage can be calibrated to coincide with those at the final stage by selecting a scale factor $a$ and by adding a hydrostatic value of pressure $b$. Only the scale factor is used for the following analysis.  }
  \label{fig:Calibration}
\end{figure}

\vspace{4mm}
\noindent {\textit{\textsf{Cross Validation}}}\\
The estimated stress can be cross-validated by comparing with that evaluated by global tissue ablation \cite{Bonnet2012}, which indicate that the stress along the medio-lateral axis of the notum increases during pupal development (Sect. \ref{sec:notum_experiment}). By applying the force-inference methods to images before ablation (the initial stage) and after the relaxation of the inner domain of cells (the final stage), stress tensors at the initial and final stages were calculated using Eq. (\ref{eq:StressTensor}) for each sample. Since the scale factor is not determined by force inference, we need to reasonably calibrate it between initial and final stages. For this, we hypothesized that the relationship between the cell pressure $P_i$ and the cell area $A_i$ is maintained between initial and final stages. In Fig. \ref{fig:Calibration}(a), estimated pressures are plotted against the cell area for initial (red) and final (blue) stages. The fitting function $P(A) = a/\sqrt{A} + b$ is given by a dimensional argument considering Laplace's law, as in the case of foam \cite{Cantat2010}.  By evaluating coefficients $a$ and $b$ by the least-squares method, we obtained the fitting curves indicated by dotted lines in Fig. \ref{fig:Calibration}(a).  Then, the scale is calibrated for these two lines (Fig. \ref{fig:Calibration}(b)).

\begin{figure}
  \centering
  \resizebox{0.5\textwidth}{!}{%
    \includegraphics{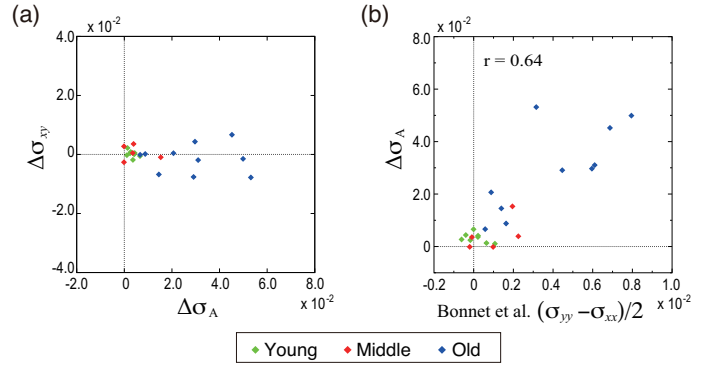}
  }
  \caption{
    Stress inference in the notum. (a) Difference of estimated stresses obtained using STP between initial (before laser ablation) and final (after the relaxation) stages ($\Delta \sigma_A = \sigma_A^f-\sigma_A^i$ and $\Delta \sigma_{xy} = \sigma_{xy}^f-\sigma_{xy}^i$).  (b) Estimated $\Delta \sigma_A$ values obtained using STP  plotted against those obtained by global ablation of tissue in \cite{Bonnet2012}.  Different colors indicate developmental stages of samples in all panels (yellow: young, red: middle, and blue: old).   }
  \label{fig:Sigmas}       
\end{figure}

Stress tensors for initial and final stages, {\boldmath $\sigma$}$^i$ and {\boldmath $\sigma$}$^f$, are estimated using STP at distinct developmental stages. We measured the normal stress difference $\sigma_{A} \equiv (\sigma_{yy}-\sigma_{xx})/2$ and $\sigma_{xy}$ in Fig. \ref{fig:Sigmas}(a), and calculated the difference of these quantities  between the initial and final stages ($\Delta \sigma_A = \sigma_A^i- \sigma_A^f$ and $\Delta \sigma_{xy} = \sigma_{xy}^i - \sigma_{xy}^f$), because they are independent of the unknown additive constant in the pressure, and characterize change of mechanical state induced by the ablation. The estimated amplitude and the difference of $\sigma_{xy}$ were smaller than those of $\sigma_{A}$ in older pupae, indicating that the stress in the notum was stronger along the $y$ axis (medio-lateral axis) (Fig. \ref{fig:Sigmas}(a)). However, $\sigma_{A}$ at young and middle stages exhibit weaker changes upon ablation of cells. These results qualitatively agree with those obtained from global ablation \cite{Bonnet2012}. To quantitatively cross-validate the two methods, $\Delta \sigma_A$ values are directly compared (Fig. \ref{fig:Sigmas}(b)). $\Delta \sigma_A$ values obtained by global ablation and STP exhibit a good correlation (with a correlation coefficient of $r = 0.64$ for calibrated data and $r = 0.59$ for uncalibrated data). By repeating the same procedures in SP and ST, we obtained $r = 0.63$ and $r = 0.59$ for calibrated and uncalibrated data using SP, and $r = 0.56$ for uncalibrated data using ST.

\subsection{Robustness to image processing error}\label{sec:Robustness}
\begin{figure}
  \centering
  \resizebox{0.48\textwidth}{!}{%
    \includegraphics{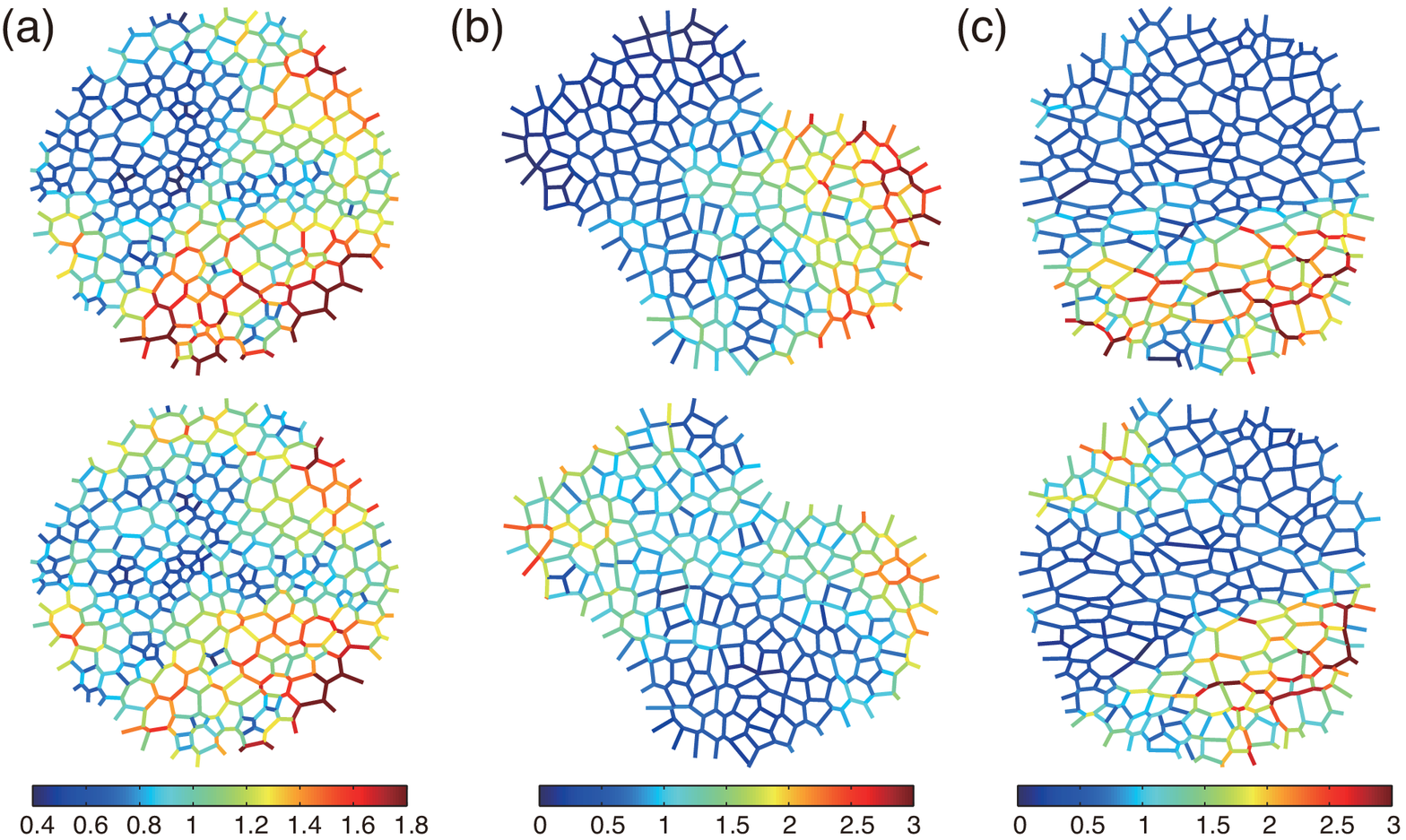}
  }
  \caption{ Examples of estimated tensions by ST for data with errors in the vertex positions. Two samples are shown for (a) simulated foam data, (b) a simulated cell population from the cell vertex model, and (c) \textit{Drosophila} notum (the method to obtain them is described in Sect. \ref{sec:Robustness}). The estimated values of tensions and the position of patches vary among samples.  }
  \label{fig:ST_robust}       
\end{figure}

We evaluated robustness of the force-inference methods to image processing error. In \cite{Ishihara2012}, a similar test for STP using wing data was briefly mentioned.  We made 100 samples by adding Gaussian noise to position coordinates of vertices in the original data, where the standard deviation of the noise is 5\% of the mean length of cell contact surfaces. Then we checked how estimated tensions and pressures deviate among the noised data. Tests were conducted for data from the foam simulation, vertex cell model, and \textit{Drosophila} notum. For reference, among the sequential images of notum taken in a very short period of time (about 1.6 seconds), the mean deviation of vertices positions extracted by image processing was less than $3\%$ of the mean contact surface lengths.

The deviations of estimated tensions among samples are large for ST; mean relative values of standard deviations (standard deviation divided by mean tension) are $17\%$, $69\%$, and $73\%$ for foam, cell vertex model, and notum, respectively. On the other hand, those for estimated tensions by STP were much smaller: $5.0\%$ (foam), $7.8\%$ (cell vertex model), and $8.5\%$ (notum). Deviations of estimated pressures for SP were $9.5 \times 10^{-4}$, $0.16$, and $0.020$ for foam, cell vertex model, and notum, and those for STP were almost the same:  $1.5 \times 10^{-3}$, $0.19$, and $0.026$. These values are sufficiently small compared to the estimated dispersion of individual cell pressures, indicating highly robust pressure estimations in SP and STP.

Moreover, we found that maps of tensions obtained by ST showed ``patches" (distinct regions where the tension seems locally uniform) and that positions of these patches differed among individual noised samples (Fig. \ref{fig:ST_robust}). These patches were also seen in tension maps estimated by ST from the original data of the wing and notum (Fig. \ref{fig:wing}(b) and Fig. \ref{fig:notum}(b)). The large error in the estimated tensions using ST can be explained by the appearance of the patches that are sensitive to errors in vertices positions.

Finally, the robustness in the stress inference was examined. The deviation among noised samples is largest in ST: for foam, the deviations of $\sigma_A$ are $1.1 \times 10^{-4}$ in ST, $5.8\times 10^{-5}$ in SP, $8.8 \times 10^{-5}$ in STP. For the cell vertex model, they were $1.0 \times 10^{-2}$, $3.6 \times 10^{-3}$, and $5.7 \times 10^{-3}$ in ST, SP, and STP, respectively. The difference between ST and STP was more significant in the experimental data of the wing (Fig. \ref{fig:wing}(a)), where the deviations of $\sigma_A$ were $1.3 \times 10^{-2}$ in ST, $5.1\times 10^{-4}$ in SP, and $1.1 \times 10^{-3}$ in STP. For the sample of notum shown in Fig. \ref{fig:notum}(a), they were $5.5 \times 10^{-3}$ in ST, $4.9\times 10^{-4}$ in SP, and $8.4 \times 10^{-4}$ in STP.

\section{Discussion} \label{discussion}
\subsection{Force-inference methods}
In the present study, we performed a comparative analysis of force and stress inference in  tissue. We employed three types of force-inference methods, which require different assumptions on the unknown variables (tension and pressure) for treating indefiniteness in the inverse problem between forces and cell shape; ST (resp.: SP) estimates only tensions (resp.: pressures) under the assumption that pressures (resp.: tensions) are uniform, and STP estimates both by having a prior that tensions are positive. We prepared four different data sets, in which the assumptions in each force-inference method are either: strictly exact, reasonable, incorrect, or not checked, which enables us to better evaluate the force-inference methods (see also Table \ref{tab:ResultSum}).

\begin{table*}
\centering
    \caption{Estimation errors of force-inference methods for artificial data.
      The errors are represented by the mean residues $\sigma^2$.}
    \label{tab:ResultSum}       
    \begin{tabular}{lccccccc}\hline \noalign{\smallskip}
      &  \multicolumn{3}{l}{Foam simulation data} && \multicolumn{3}{l}{Cell vertex simulation data}    \\
      estimation error   & ~ST & ~SP & STP && ~ST & ~SP & STP \\ \hline \noalign{\smallskip}
      tension   & $4.5 \times 10^{-2}$  & --- & $3.4 \times 10^{-4}$  &&
      $1.7 \times 10^{-1} $& --- & $3.8 \times 10^{-3}$\\
      pressure   & ---& $3.2 \times 10^{-7}$  & $3.4 \times 10^{-7}$ &&
      --- & $4.7 \times 10^{-3}$ & $1.0 \times 10^{-3}$
      \\  \noalign{\smallskip} \hline
    \end{tabular}
\end{table*}

The assumption in SP that all tensions are uniform is strictly exact in foam and is not exact in the cell vertex model or in tissue (and the variance of tensions may differ in each tissue).  Estimates of pressures exhibited high accuracy in both  the numerical data from foam and  the cell vertex model (Table \ref{tab:ResultSum}), suggesting that the difference in the cell area is a good indication of cell pressures in the artificial data employed in this study. Whether SP provides reasonable estimates of pressures in tissues where tensions vary greatly among contact surfaces is a subject for future study.

The assumption in ST that all pressures are uniform is incorrect in foam and in the cell vertex model and is not directly checked in tissue. The errors of tension inference obtained using ST are large in both the numerical data from foam and the cell vertex model. In foam, the estimated tensions exhibit a positive correlation with the length of the contact surface, which disagrees with the physical nature of foam. In the cell vertex model, estimated tensions are significantly more disperse than the true ones.  

The large estimation error in ST may result from its incorrect assumption. Another, but not mutually exclusive, possibility is indicated by our results: adding noise in input data causes the appearance of "patches" in a map of tensions. Note that the force-balance equations (\ref{eq:Force_Vertex_X}) and (\ref{eq:Force_Vertex_Y}) provide only local information and that the force-balance equations become non-exact upon addition of error in the vertex positions. Thus, one can speculate that a gradual spatial change in tensions and pressures (\textit{i.e.,} modes with long wavelength) can be allowed, which generates a spatial modulation comparable to system-size. The patches of tensions generated by long-wavelength modes are sensitive to image processing error as indicated by our observation that the positions of the patches vary among noised samples (Fig. \ref{fig:ST_robust}).  The positions of the vertices and the force-balance equations are exact in the original artificial data, while image processing errors are unavoidable in experimental data. This is the reason why the patches of tensions were much more evident in the experimental data than in the artificial ones (compare Fig. \ref{fig:Foam}(b) and Fig. \ref{fig:CVM}(b) with Fig. \ref{fig:wing}(b) and Fig. \ref{fig:notum}(b)). In contrast, the force inference by STP is more accurate and more robust to image processing error. In fact, in STP, the prior that expects $T_{ij}$ should be close to $T_0>0$ works as the `regularization term' to avoid over-fitting by eliminating the long-wavelength mode, which makes the force inference more robust (see also Sect. \ref{sec:Variable-reductionandBayes}).

The assumption in STP that tensions are distributed around a positive value is reasonable in epithelial tissue, as suggested by laser ablation of individual contact surfaces \cite{Rauzi2008,Hutson2003}. Our results clearly indicate that estimates of pressures and tensions correlate well to the true values in both the numerical data from foam and the cell vertex model. To summarize, STP gives better estimates of tensions than ST. SP is slightly better at pressure inference than STP for foam data (Fig. \ref{fig:Foam}(f)), in which the assumption $T_{ij} = 1$ is the true physical nature of the system, whereas STP gives slightly better results than SP for the data generated by the cell vertex model, in which tensions are not uniform (Fig. \ref{fig:CVM}(f)).

The application of force-inference methods to experimental data yielded results consistent with those from artificial data. Estimates of pressures by SP and STP are highly correlated in the \textit{Drosophila} wing and notum. In contrast, estimates of tensions by ST vary more widely than those by STP; several contact surfaces are estimated to have almost zero tension, which may result from the spatial distribution of tensions where the strength of tensions are significantly different among neighboring patches.

\subsection{Summary of the cross-validation}
We compared two different stress measurement methods in  tissue: One  is robust and model-independent \cite{Bonnet2012} and one is non-invasive and can yield a space-time map (\textit{i.e.}, STP) \cite{Ishihara2012}. Annular ablation of  tissue showed that the stress of the notum increases along the medio-lateral axis during  pupal development. STP can detect such developmental changes in stress. Moreover, the value of $\Delta \sigma_A$ inferred using STP better agrees with experimental data than those found using ST and SP. These results serve as double-checks for the anisotropy of stress in the notum and in turn this reinforces the validation of the annular ablation method of stress measurement.

\subsection{The curvature of a contact surface} \label{sec:curvature_discussion}
In the present study, we neglected the curvature of the cell contact surface and obtained good estimates of pressures in SP and STP. Note that neglecting the curvature does not imply that the difference in pressures among cells is ignored. It was previously discussed that the error of stress evaluation under the straight-edge approximation could be small, if the curvature is small \cite{Janiaud2005}. Neglecting the curvature was also demonstrated to give a good estimation \textit{in vivo} by checking Laplace's law for the tension estimated under the straight-edge approximation \cite{Chiou2012}. One advantage of neglecting the curvature of the contact surface is that we can bypass having to make a precise curvature measurement from an image, which is known to be difficult and/or error-prone \cite{Kovalevsky2001}. If the pressure difference between cells is large, the curvature cannot be neglected; hence the force-inference method needs to be extended to use information about the relationship between forces and the curvature.

\subsection{A practical choice of force and stress inference}
Selection of the most suitable force-inference method is dependent on the nature of the system and experimental design. As summarized above (Table \ref{tab:ResultSum}), tension inference obtained using STP is better than that obtained using ST in accuracy and in robustness and STP gives more robust stress inference than ST, whereas pressure and stress inference obtained using STP and SP are comparable.  On the other hand, SP and ST involve solving a linear equation only once, STP requires performing QR decomposition many times during the maximization process \cite{Ishihara2012}. Although it requires less than $\sim \negthickspace10$ min for most of the data investigated in this study ($\sim \negthickspace200$ cells), computational time increases for data with a larger number of  cells. Collectively, STP should be the first choice for tension, pressure, and stress inference. SP can be useful for stress and pressure inference, when the variance of tensions are known to be sufficiently small as in the notum, and when there are large numbers of cells (\textit{e.g.}, more than thousands) in the system of interest.

In conclusion, the present study strengthens the validity of our force-inference \cite{Ishihara2012} and stress measurement \cite{Bonnet2012} methods. Their future improvement would accelerate studies of the physical regulations of development.

\vspace{1em} \textbf{Acknowledgements}
We thank Philippe Marcq for fruitful discussions and Floris Bosveld for the flies and discussions. We are also grateful to the RIKEN 
BSI-Olympus Collaboration Center for the imaging equipment. This work was supported by PRESTO JST (S.I.), the RIKEN SPDR 
and Incentive Research Grant programs, and MEXT (K.S.).

\end{document}